\documentclass[pra,tightenlines,showpacs,nofootinbib]{revtex4}

\usepackage{dcolumn,bm,graphicx,amsmath}
\usepackage
{hyperref}

\begin{document}
\preprint{UNR June 2005-\today }
\title{ Revised Huang-Yang multipolar pseudopotential }

\author{Andrei Derevianko}
\email{andrei@unr.edu}
\affiliation{Department of Physics, University of Nevada, Reno, Nevada 89557}

\date{\today}

\begin{abstract}
A number of authors have recently pointed out inconsistencies of
results obtained with the Huang-Yang multipolar pseudo-potential for low-energy
scattering [K. Huang and K. C. Yang, Phys. Rev. A {\bf 105}, 767 (1957);
later revised in K. Huang, ``Statistical Mechanics'', (Wiley, New York, 1963)].
The {\em conceptual} validity of their original derivation has been questioned.
Here I show that these inconsistencies are rather due to an {\em algebraic}
mistake made by Huang and Yang.  With the corrected error,
 I present the revised version of the multipolar pseudo-potential.
\end{abstract}


\maketitle

\section{Introduction}
Most many-body problems require evaluating matrix elements of interparticle interaction in
the plane-wave basis. For a typical interaction with a short-range repulsive ``hard'' core
such integrals diverge. Nevertheless,  this problem can be made tractable with so-called
pseudo-potential technique, which is usually attributed to Fermi~\cite{Fer36}. In this technique,
a suitably chosen pseudo-potential replaces the true interaction; the solution of
the Schrodinger equation with the pseudo-potential must reproduce the
long-range behavior of the wave-function derived with the original interaction. Since the most
fundamental
ingredient of the quantum mechanical treatment, the
wavefunction, is properly recovered, any  non-singular property should be well
approximated in this approach.

In the Huang-Yang (HY) construction~\cite{HuaYan57,Hua63},
the pseudo-potential $V_{\mathrm{ps}}$ is determined as a multipolar
expansion over delta-function (``lumped'') contributions.
For a partial wave expansion,
$\psi \left( \mathbf{r}\right) =\sum_{l,m}\psi _{lm}\left( r\right) Y_{lm}\left( \hat{r}\right) $,
\begin{eqnarray}
V_{\mathrm{ps}}\psi\left(  \mathbf{r}\right)    & = &
\sum
_{l,m}Y_{lm}\left(  \hat{r}\right)  ~\hat{v}_{l}~\psi_{lm}\left(  r\right) \, , \label{Eq:VPS}\\
\hat{v}_{l}~\psi_{lm}\left(  r\right)    & = & -f_{l}\frac{\hbar^{2}}{2\mu} \frac{\delta\left(
r\right)  }{r^{l+2}}\frac{\tan\eta_{l}}{k^{2l+1}}\left(  \frac{\partial
}{\partial r}\right)  ^{2l+1}\left(  r^{l+1}\psi_{lm}
\right)   \, ,\nonumber
\end{eqnarray}
where $\mu$ is the reduced mass of the interacting pair, $k$ is the conventional collision
wavevector, and $\eta_l$ is the phase shift for a partial wave $l$. As to the prefactor $f_l$,
the values of the prefactor differ in the original HY
paper~\cite{HuaYan57} and in the  Huang's textbook~\cite{Hua63}, published later.
 Namely the value of the prefactor
from the Huang's textbook is used in the literature.
Here, by tracing steps in the Huang  derivation, I point out that still there  remains
an algebraic mistake
in the value of the pre-factor $f_l$. The new revised value of the prefactor is
\begin{equation}
f_l^\mathrm{revised} =  \frac{2l+1}{l+1} f_l^\mathrm{Huang} =  \frac{(2l+1)!!}{l!2^{l}} \, ,
\label{Eq:flRevised}
\end{equation}
where $f_l^\mathrm{Huang}$ is the original (erroneous) pre-factor from Ref.\cite{Hua63}.

Notice that the $s$-wave contribution ($l=0$) to $V_{\mathrm{ps}}$ is not affected
by the correction. Since the
$s$ waves dominate low-energy collision physics, certain inconsistencies
for higher partial waves has not been noticed
until very recently~\cite{RotFel01,StoSilBol05}, when
higher order multipoles  became a subject of interest.
For example, for identical fermions the $s$-wave contribution vanishes because
of the symmetry arguments,
and one has to consider  $p$-wave scattering in particular.
There are other scenarios, e.g., a resonant coupling of $d$-waves, when
the multipoles beyond $l=0$ become relevant. Also the strong coupling
of higher partial waves to  $s$-waves is a prominent feature for
anisotropic (e.g., dipolar) interactions.

First, \citet{RotFel01} have considered mean-field correction to energies of
trapped fermions. They have derived their own version of the pseudo-potential aimed at
reproducing the energy shifts. By computing the same corrections
with the HY pseudo-potential, these authors
found that each multipolar contribution in the HY pseudo-potential must be multiplied
by a factor of $(2l+1)/(l+1)$ (just as in Eq.(\ref{Eq:flRevised}))
to bring the computed correction into an agreement with their independent results.
These authors concluded that the HY pseudo potential
``is not a proper effective interaction for a mean field description of dilute
quantum gases that goes beyond $s$-wave interactions''. Unfortunately,
this statement can be interpreted as if two different
versions of the pseudopotential were to be employed: one version for continuum
and another version for bound-state problems.

Second, an alternative derivation of the pseudopotential has been
presented by \citet{StoSilBol05}. Instead of delta-function lumped at the origin,
these authors have proposed to use a shell pseudo-potential, i.e., delta-function
placed at the surface of spherical shell.
In the limit of zero radius of the shell each multipolar
term of the HY pseudo-potential is obtained. As in Ref.~\cite{RotFel01} this limit
produced multipolar terms differing by a
factor of $(2l+1)/(l+1)$ from the HY terms. Because of these missing factors, the
authors of Ref.~\cite{StoSilBol05} claimed that there is
a  ``fundamental problem'' with the Huang-Yang derivation.

Here I demonstrate
that there is no {\em conceptual} problem with the Huang-Yang construction.
Rather there is an {\em algebraic} mistake in the derivation~\cite{Hua63}. I have traced  the
error to the  erroneous application
of the Gauss theorem in~\cite{Hua63}.
With the corrected derivation, this additional  factor of $(2l+1)/(l+1)$ is fully recovered.
There is no need to introduce the intermediate $\delta$-shell pseudo-potential as proposed in
Ref.~\cite{StoSilBol05}. In other words, there is a reconciliation of the seminal paper by
Huang and Yang
with the more recent derivations.

This short paper is organized as follows. In Section~\ref{Sec:Born}, by deriving phase shifts in the
Born approximation
with the HY pseudo-potential I present another demonstration that there is a consistency problem with
the original HY expressions. In Section~\ref{Sec:Err}, I will point out the algebraic error in
Huang-Yang derivation.

\section{Integral equation for the phase shifts and Huang-Yang pseudo-potential}
\label{Sec:Born}
As discussed in the introduction there is some evidence from the literature~\cite{RotFel01,StoSilBol05}
that there is a difficulty with the original HY formula.
Below I provide an alternative self-consistency check of the HY pseudopotential based on the
integral equation for the phase shifts. I  arrive at the value of the prefactor which
indeed differs from the one prescribed by Huang.

The radial Lippmann-Schwinger equation for a finite-ranged spherically-symmetric potential leads
to the following implicit equation for the phase shifts~\cite{Fri04}
\begin{equation}
\tan\eta_{l}=-\sqrt{\frac{2\mu~}{\hbar^{2}}\frac{\pi}{k}}\int_{0}^{\infty
}kr~j_{l}\left(  kr\right)  V\left(  r\right)  \phi_{l}\left(  r\right)  dr \, .
\label{Eq:LStanShift}
\end{equation}
Here $\phi_{l}\left(  r\right)$ is the properly normalized exact solution of the scattering
problem. For the HY pseudo-potential this solution is by construction
\begin{equation}
\phi_{l}\left(  r\right)  =\sqrt{\frac{2\mu~}{\hbar^{2}}\frac{1}{\pi k}%
}\left\{  krj_{l}\left(  kr\right)  -\tan\eta_{l}~kr~n_{l}\left(  kr\right)
\right\} \, , \label{Eq:Sol}
\end{equation}
with $j_l(kr)$ and $n_l(kr)$  being the conventional spherical Bessel and Neumann functions. At small
values of the argument,
\[
j_{l}\left(  z\right)  \simeq\frac{1}{\left(  2l+1\right)  !!}z^{l}%
, \, n_{l}\left(  z\right)  \simeq-\frac{\left(  2l-1\right)  !!}{z^{l+1}} \, .
\]
Now we substitute the solution~(\ref{Eq:Sol}) into the Eq.(\ref{Eq:LStanShift}) with
the HY pseudo-potential, Eq.(\ref{Eq:VPS}). The part of the solution proportional to the Neumann
function vanishes upon differentiation and an intermediate result is
\begin{eqnarray*}
\tan\eta_{l}&=&f_{l}\frac{1}{\left[  \left(  2l+1\right)  !!\right]  ^{2}}%
\tan\eta_{l}\frac{1}{k^{2l+2}} \\
& &\int_{0}^{\infty}\left(  kr\right)  ^{l+1}%
\frac{\delta\left(  r\right)  }{r^{l+2}}\left(  \frac{\partial}{\partial
r}\right)  ^{2l+1}\left\{  r^{l+1}\left(  kr\right)  ^{l+1}\right\}  dr \, .
\end{eqnarray*}
This equation allows us to obtain the value of the prefactor $f_l$.
Taking into account $\left(  \frac{\partial}{\partial r}\right)
^{2l+1}r^{2l+2}=\left(  2l+1\right)! \, r$, one arrives at the value of prefactor
$
f_{l}  = \left(  2l+1\right)  !! /(2^{l}l!) \, ,
$
i.e., differing from the Huang's prefactor by $(2l+1)/(l+1)$.
Again, as in Refs.~\cite{RotFel01,StoSilBol05}, we conclude that there is
a self-consistency problem with the HY formula for $l>0$ multipoles. Below, I will trace the steps
in Huang's derivation and I will point out the mistake in his derivation.

\section{Tracing the algebraic error in Huang-Yang derivation}
\label{Sec:Err}
While the original pseudopotential has been introduced in the Huang-Yang paper~\cite{HuaYan57},
there are certain mistakes in the final equations. These
formulae have been revised later by Huang in his textbook~\cite{Hua63}.
Also additional details of the derivation are given there. Tracing the
steps in the derivation, I found the error in the chain of
equations (B.13) of Ref.~\cite{Hua63}. Here Huang integrates over a small spherical volume
$V_{\varepsilon}$ of infinitesimal radius $\varepsilon$
\begin{align*}
I_{\varepsilon} &  =\int_{V_{\varepsilon}}d^{3}r\text{ }r^{l}\left\{  \frac
{1}{r^{2}}\frac{d}{dr}r^{2}\frac{d}{dr}n_{l}\left(  kr\right)  -\frac{l\left(
l+1\right)  }{r^{2}}n_{l}\left(  kr\right)  \right\} \, .
\end{align*}
Noticing that $
\nabla^{2}r^{l}=\frac{l\left(  l+1\right)  }{r^{2}} r^l $
and that for a spherically-symmetric function%
\[
\nabla^{2}f\left(  r\right)  =\frac{1}{r^{2}}\frac{d}{dr}r^{2}\frac{d}%
{dr}f\left(  r\right) \, ,
\]
this expression is brought into a form suitable for application of the
Green's theorem (the Green's second identity),
\[
I_{\varepsilon} = \int_{V_{\varepsilon}}d^{3}r\text{ }\left\{  r^{l}\nabla^{2}n_{l}\left(
kr\right)  -n_{l}\left(  kr\right)  \nabla^{2}r^{l}\right\} \,.
\]
The second Green's identity reads~\cite{Kap91}
\[
\int_{V} d^3r \, \left(  \phi\mathbf{\nabla}^{2}\psi-\psi\mathbf{\nabla}^{2}%
\phi\right) =\oint_{S}d\mathbf{S}\cdot\left(  \phi\mathbf{\nabla
}\psi-\psi\mathbf{\nabla}\phi\right)  \, ,
\]%
leading to
\[
I_{\varepsilon}=\oint_{S}d\mathbf{S}\cdot\left(  r^{l}\mathbf{\nabla}%
n_{l}\left(  kr\right)  -n_{l}\left(  kr\right)  \mathbf{\nabla}r^{l}\right) \, ,
\]
while Huang's formula reads%
\[
I_{\varepsilon}^{\text{Huang}}=\oint_{S}d\mathbf{S}\cdot\left(  r^{l}%
\mathbf{\nabla}n_{l}\left(  kr\right)  \right) \, .
\]
The error is here. Huang is missing the second term, $-n_{l}\left(  kr\right)
\mathbf{\nabla}r^{l}$. Continuing the chain of equations, we obtain
\begin{align*}
I_{\varepsilon} &  =\oint_{S}\left(  r^{l}\frac{d}{dr}n_{l}\left(  kr\right)
-n_{l}\left(  kr\right)  \frac{d}{dr}r^{l}\right)  dS=\\
&  =\frac{\left(  2l-1\right)  !!}{k^{l+1}}\oint_{S}\left(  -r^{l}\frac{d}%
{dr}\frac{1}{r^{l+1}}+\frac{1}{r^{l+1}}\frac{d}{dr}r^{l}\right)  dS=\\
&  \frac{\left(  2l-1\right)  !!}{k^{l+1}}\oint_{S}\left(  \left(  l+1\right)
\frac{1}{r^{2}}+l\frac{1}{r^{2}}\right)  dS \, .
\end{align*}
Namely the second  (missed term) provides the additional contribution of $l$.
Finally,%
\[
I_{\varepsilon}=4\pi\frac{\left(  2l-1\right)  !!}{k^{l+1}}\left(
2l+1\right)
\]
or Eq.(B.14) of Ref.~\cite{Hua63} should read%
\[
F_{l}\left(  r\right)  =\frac{\left(  2l-1\right)  !!}{k^{l+1}}\left(
2l+1\right)  \frac{\delta\left(  r\right)  }{r^{l+2}} \, .
\]
Combining this result with the rest of the derivation we arrive at
the revised Huang-Yang pseudopotential Eqs.~(\ref{Eq:VPS},\ref{Eq:flRevised}).

To summarize, here a revised form  of the low-energy multipolar pseudopotential by Huang and Yang\cite{HuaYan57,Hua63} has
 been presented.
A mistake in the original derivation has been pointed out. The present paper
reconciles the seminal Huang-Yang construction with more recent results from the literature.

This
work was supported in part by the  NSF Grant No. PHY-0354876.



\end{document}